\documentclass[aps,prb,amssymb,noshowpacs,twocolumn,tightenlines]{revtex4}
\usepackage{graphicx}
\usepackage{amsmath,bm}
\usepackage[mathscr]{euscript}
\usepackage{dcolumn}

\begin{document}
\title{On-Top Density in the Nonlinear Metallic Screening \\
and its Implication on the Exchange-Correlation Energy Functional}

\author{Yasutami Takada}
\thanks{Email: takada@issp.u-tokyo.ac.jp; published in Eur. Phys. J. B; 
https://doi.1140/epjb/e2018-90111-0}
\affiliation{Institute for Solid State Physics, University of Tokyo,
             Kashiwa, Chiba 277-8581, Japan}


\begin{abstract}
In comparison with the accurate data on the on-top electron density $n(0)$ in the 
proton-embedded electron gas with the density parameter $r_s$ in the range $1-12$ 
obtained by diffusion Monte Carlo (DMC) simulations, we have successfully 
constructed an alternative form of the exchange-correlation energy functional 
in the density functional theory by imposing the constraint due to the cusp theorem 
on the well-known Perdew-Burke-Ernzerhof (PBE) functional. Although PBE does not, 
our functional, referred to as the cusp-corrected PBE (ccPBE), reproduces 
the DMC data on $n(0)$ in the entire range of $r_s$. 
\end{abstract}


\maketitle


\section{Introduction}
\label{Sec_1}

An atom, especially hydrogen, immersed into the otherwise homogeneous electron 
gas (EG) has been investigated for more than four decades not only in the density 
functional theory (DFT), mostly in its local-density approximation (LDA)~\cite{PS74,ABPS76,ZSSR77,JSN78,BM78,B79,JFE79,Norskov79,SZ80,PNM81,PNM83,PNM91,Song04,NKT05}, 
but also in various forms of many-body theories~\cite{PSCP76,JS78,GJS78,JGS78,AP79,GS85,SE91}, including diffusion Monte Carlo (DMC)~\cite{STA88} and variational 
Monte Carlo (VMC)~\cite{DA07} simulations. The primary motivation of those studies 
is to construct an appropriate theory for the nonlinear response of metallic 
electrons to an impurity point charge $+Ze$, but the basic physical concept with 
which they were concerned remains the same as that in the linear-response theory, 
known as Thomas-Fermi (TF)~\cite{Thomas27,Fermi27} (or Debye and H\"uckel~\cite{DH23}) 
screening of the impurity charge with a short screening length $\lambda_{\rm TF}$ 
which is about the same as $k_{\rm F}^{-1}$, where $k_{\rm F}$ is the Fermi wave 
number of EG. 

Recently, by studying a proton (the case of $Z=1$) embedded in EG with use of both 
LDA and DMC, the present author has gained a new insight into this problem~\cite{TMY15}; 
the concept of Kondo screening of the spin of hydrogen with a long screening length 
$\lambda_{\rm K}\ (\gg k_{\rm F}^{-1})$~\cite{Hewson93} is found to be relevant to 
this system and a sharp transition from TF to Kondo screening is shown to exist with 
the decrease of the metallic electron density $n_0\ (=k_{\rm F}^3/3\pi^2)$ from the 
high-density limit. At the same time, the results in DMC are found to be well 
approximated by those in LDA in the density region characterized by Kondo screening 
because of the slowly-varying nature of the electron density distribution $n({\bm r})$ 
around the proton due to the long $\lambda_{\rm K}$. 

In the high-density region characterized by TF screening, on the other hand, 
a relatively large difference can be seen in $n({\bm r})$ between DMC and LDA. In 
particular, an unexpected feature of $n({\bm r})$ is found at the proton position 
or the on-top density $n(0)$; in DMC, $n_{\rm DMC}(0)$, is lower than that in LDA, 
$n_{\rm LDA}(0)$, at high densities, namely, $r_s<1.66$ with the conventional 
density parameter $r_s \equiv (\alpha k_{\rm F}a_{\rm B})^{-1}$, while the opposite 
is the case for $r_s>1.66$. Here we define $\alpha=(4/9\pi)^{1/3} \approx 0.5211$ 
and $a_{\rm B}$ is the Bohr radius. (We will use atomic units hereafter.) 

According to a physical argument~\cite{KO79}, we obtain larger $n(0)$ for stronger 
exchange-correlation (xc) effect, implying that as long as we believe that 
$n_{\rm DMC}(0)$ is sufficiently accurate, LDA is found to provide a too strong 
xc effect for $r_s<1.66$ but a too weak one for $r_s>1.66$. This interesting 
crossover behavior with the increase of $r_s$ in describing the xc effect in LDA 
has never been known. 


Because $n({\bm r})$ varies very weakly even for $r< \lambda_{\rm TF}$ in densely 
packed systems such as the high-density EG for $r_s<1.66$, it is natural to expect 
that a small density-gradient correction to LDA will be enough to obtain a result 
of $n(0)$ much better than LDA, but actually the situation becomes worse in the 
generalized gradient approximation (GGA) in the Perdew-Burke-Ernzerhof (PBE) 
version~\cite{PBE}; namely, for $r_s<1.66$, the difference of PBE from DMC becomes 
larger than that of LDA. One might imagine that not PBE but the {\it accurate} 
gradient expansion~\cite{AK85} as included in PBEsol~\cite{PBEsol} is needed to 
obtain better $n(0)$, but this is not the case; no improvement on LDA is achieved 
even in PBEsol. Thus we come to notice that it is a nontrivial work to reproduce 
$n_{\rm DMC}(0)$ for the case of $r_s<1.66$ in the framework of GGA. 


In pursuit of a key ingredient to improve on PBE in the present problem 
with retaining exact conditions which make PBE reliable, as listed in Table I in 
Ref.~\cite{SSTP04}, we come across the importance of the cusp 
theorem~\cite{Kato57,CA82,PS03} which dictates that $n({\bm r})$ near the impurity 
atom behaves rigorously in such a manner as
\begin{align}
n(\bm{r}) \xrightarrow[r\, \approx\, 0]{}n_{\rm cusp}(r)\equiv n(0)\exp(-2Zr),
\label{eq:1}
\end{align}
so as to make a compromise with the singular Coulomb potential term $-Z/|{\bm r}|$. 
Although $n_{\rm LDA}({\bm r})$ satisfies Eq.~(\ref{eq:1}), $n({\bm r})$ in 
PBE or PBEsol does not, indicating that the worse performance of PBE/PBEsol 
in determining $n(0)$ might originate from the violation of the cusp theorem. 


Generally it is not believed that we can make the cusp theorem obeyed in the 
framework of GGA~\cite{SSTP04,PRTSSC05} and it is usually thought that some 
form of meta-GGA~\cite{TPSS03,PRCCS09,SRP15} is needed to satisfy it. Therefore the 
inclusion of the cusp theorem into a GGA-based scheme is really a challenge. 
In this paper we take up this challenge and set the goal of this paper in the 
following way; we just try to modify the spin-resolved xc energy functional 
$E^{\rm xc}[n_{\sigma}]$ in PBE by imposing the constraint due to the cusp theorem 
in addition to the exact conditions already obeyed by PBE and then we tune up some 
free parameters involved in the modified $E^{\rm xc}[n_{\sigma}]$ so as to reproduce 
$n_{\rm DMC}(0)$ in the wide range of $r_s$, i.e., $1 \le r_s \le 12$ 
where the DMC data are available. 

We will leave a comprehensive test of this modified $E^{\rm xc}[n_{\sigma}]$ 
(which will be referred to as ``cusp-corrected'' PBE or ccPBE) for a variety of 
real materials for the future, but because ccPBE provides the different results of 
$n(0)$ from those in PBE only for $r_s<1.66$, ccPBE and PBE will give, more or less, 
similar results for almost all real materials. One important exception is the solid 
hydrogen under very high pressures~\cite{AF13,CHCM16} in which $1.1<r_s<1.7$. 
Thus ccPBE may be expected to be useful only for solid hydrogen. 

This paper is organized as follows: In Sect.~\ref{sec:2}, we introduce the sytem 
to be treated, explain the calculation methods, and account for the issues arisen 
from the data calculated on $n(0)$. In Sect.~\ref{sec:3}, we construct ccPBE and 
give the calculated results for $n(0)$ in ccPBE in comparison with those in DMC. 
Finally in Sect.~\ref{sec:4}, we give a summary of this paper and make several comments.

\section{Atom embedded in the jellium sphere}
\label{sec:2}
\subsection{Hamiltonian}
\label{sec:2-1}

Because Monte Carlo simulations can treat only a finite number of electrons, let us 
consider not bulk jellium but a jellium sphere of radius $R$ 
and average density $n_0$ and then put a neutral atom of atomic number $Z$ at 
${\bm r}\!=\!{\bm 0}$ (the center of the sphere). The number of electrons in the 
jellium sphere is $4\pi R^3n_0/3\!=\!(R/r_s)^3$, so that the total electron number 
$N$ is equal to $Z\!+\!(R/r_s)^3$, satisfying global neutrality, from which we 
obtain $R=(N\!-\!Z)^{1/3}r_s$. The Hamiltonian $H$ for electrons in this system 
is given as
\begin{align}
H \!=\! -\sum_i \! \frac{\mbox{\boldmath$\nabla$}_i^2}{2} \!+\! \frac{1}{2} 
\sum_{i \neq j} \frac{1}{|{\bm{r}_{i}}\! -\! {\bm{r}_{j}}|}  
\!+\! \sum_{i} v_{\rm ext}({\bm{r}_{i}}),
\label{eq:2}
\end{align}
where the external potential working on an electron $v_{\rm ext}({\bm r})$ is 
composed of the potential from the nucleus and that from the positive background, 
written as
\begin{align}
v_{\rm ext}({\bm r})=& -\frac{Z}{r}
-\frac{N-Z}{2}\,\frac{3R^2-r^2}{R^3}\,\theta(R-r)
\nonumber \\
&-\frac{N-Z}{r}\,\theta(r-R),
\label{eq:3}
\end{align}
with $r=|{\bm r}|$ and $\theta(x)$ the Heaviside step function. In solving 
Eq.~(\ref{eq:2}), we impose the fixed boundary condition to make the wave function 
vanish at $|{\bm r}_i|\!=\!R$. From a computational point of view, this boundary 
condition is indispensable to obtain rapidly and stably convergent results 
in the closed-shell condition. 

\subsection{DFT and the Kohn-Sham scheme}
\label{sec:2-2}

In DFT, the spin-resolved ground-state density $n_{\sigma}({\bm r})$ for $H$ 
in Eq.~(\ref{eq:2}) is rigorously determined by the map to a noninteracting 
reference system which is solved by the Kohn-Sham (KS) equation, written as
\begin{equation}
\left [ -\mbox{\boldmath$\nabla$}^2/2 + v^{\rm KS}_{\sigma}(\bm{r}) 
\right ] \phi_{i\sigma} (\bm{r}) = \varepsilon_{i\sigma}\phi_{i\sigma}(\bm{r}),
\label{eq:4}
\end{equation}
where $\varepsilon_{i\sigma}$ and $\phi_{i\sigma}$ are the energy level and 
the normalized wave function for KS orbital $i$ and spin $\sigma$, respectively, 
and $v^{\rm KS}_{\sigma}(\bm{r})$ is the KS potential, given by 
\begin{align}
v^{\rm KS}_{\sigma}(\bm{r})\! =\! v_{\rm ext}({\bm r})\!+\!
\int \! d\bm{r}'\, \frac{n(\bm{r}')}{|\bm{r}\! - \!\bm{r}'|}
\!+\!v^{\rm xc}_{\sigma} ({\bm r};[n_{\sigma}]),
\label{eq:5}
\end{align}
where $v^{\rm xc}_{\sigma} (\bm{r};[n_{\sigma}])$ is derived from 
$E^{\rm xc}[n_{\sigma}]$ through the functional derivative as
\begin{equation}
v^{\rm xc}_{\sigma} (\bm{r};[n_{\sigma}])= 
\delta E^{\rm xc}[n_{\sigma}]/\delta n_{\sigma}(\bm{r}).
\label{eq:6}
\end{equation}
With use of the lowest-$N_{\sigma}$ KS orbitals, $n_{\sigma}(\bm{r})$ is given by 
\begin{equation}
n_{\sigma}(\bm{r}) = \sum_{i=1}^{N_{\sigma}} |\phi_{i\sigma} (\bm{r})|^2, 
\label{eq:7}
\end{equation}
and $n(\bm{r})$ is the sum of $n_{\uparrow}(\bm{r})$ and 
$n_{\downarrow}(\bm{r})$. The spin density $n_{\sigma}(\bm{r})$ and 
consequently $N_{\sigma}$ with $N=\sum_{\sigma}N_{\sigma}$ should be determined 
by the self-consistent solution of Eqs.~(\ref{eq:4})-(\ref{eq:7}), together 
with the fixed boundary condition 
\begin{equation}
\phi_{i\sigma} (\bm{r})=0,
\label{eq:8}
\end{equation}
at $r\!=R\!=\!(N\!-\!Z)^{1/3}r_s$. This boundary condition is imposed 
to make a direct comparison of the results in DFT-based schemes with those in DMC.

\subsection{LSDA}
\label{sec:2-3}

In order to implement the above KS scheme, we need to know some concrete form of 
$E^{\rm xc}[n_{\sigma}({\bm r})]$. In the local-spin density approximation (LSDA), 
it is given by
\begin{align}
E^{\rm xc}[n_{\sigma}({\bm r})]=\int d{\bm r}\, n({\bm r}) 
\epsilon_{\rm xc}^{\rm unif}\bigl (r_s({\bm r}),\zeta({\bm r})\bigr ),
\label{eq:9}
\end{align} 
where $n({\bm r})\!=\!n_{\uparrow}({\bm r})\!+\!n_{\downarrow}({\bm r})$ and 
$\epsilon_{\rm xc}^{\rm unif}(r_s,\zeta)$ is the xc energy per electron 
for the homogeneous electron gas with the density parameter $r_s=(3/4\pi n)^{1/3}$ 
and the spin polarization $\zeta\!=\!(n_{\uparrow}\!-\!n_{\downarrow})/n$. 
Usually, $\epsilon_{\rm xc}^{\rm unif}(r_s,\zeta)$ is divided into two parts; 
the exchange part $\epsilon_{\rm x}^{\rm unif}(r_s,\zeta)$ and the correlation part 
$\epsilon_{\rm c}^{\rm unif}(r_s,\zeta)$, both of which are concretely given in 
Ref.~\cite{PW92}, but we can simply write $\epsilon_{\rm x}^{\rm unif}(r_s,\zeta)$ as 
\begin{align}
\epsilon_{\rm x}^{\rm unif}(r_s,\zeta)=\epsilon_{\rm x}^{\rm unif}(r_s)
\frac{(1+\zeta)^{4/3}+(1-\zeta)^{4/3}}{2}, 
\label{eq:10}
\end{align} 
with $\epsilon_{\rm x}^{\rm unif}(r_s)=-(3/4)(3/2\pi)^{2/3}/r_s$.

\subsection{PBE}
\label{sec:2-4}

In GGA, $E^{\rm xc}[n_{\sigma}({\bm r})]$ is given as a functional of not only 
$n_{\sigma}({\bm r})$ but also its first derivative 
$\mbox{\boldmath$\nabla$}n_{\sigma}({\bm r})$. In its PBE version, 
$E^{\rm xc}[n_{\sigma}({\bm r})]$ is assumed to be 
\begin{align}
E^{\rm xc}[n_{\sigma}({\bm r})]=\frac{E^{\rm x}[2n_{\uparrow}({\bm r})]
\!+\!E^{\rm x}[2n_{\downarrow}({\bm r})]}{2}\!+\!E^{\rm c}[n_{\sigma}({\bm r})],
\label{eq:11}
\end{align} 
with the exchange energy functional $E^{\rm x}[n({\bm r})]$, written as 
\begin{align}
E^{\rm x}[n({\bm r})]=\int d{\bm r}\, n({\bm r}) 
\epsilon_{\rm x}^{\rm unif}(r_s)F_{\rm x}(s),
\label{eq:12}
\end{align} 
where $s=s({\bm r})$ is the normalized derivative, defined by 
\begin{align}
s({\bm r}) = \frac{|\mbox{\boldmath$\nabla$}n({\bm r})|}
{2k_F({\bm r)}n({\bm r})}
= \frac{|\mbox{\boldmath$\nabla$}n({\bm r})|}
{2(3\pi^2)^{1/3}n({\bm r})^{4/3}},
\label{eq:13}
\end{align} 
with $k_F({\bm r})=[3\pi^2n({\bm r})]^{1/3}$ and $F_{\rm x}(s)$ is given by
\begin{align}
F_{\rm x}(s)=1+\kappa-\frac{\kappa}{1+\mu_{\rm PBE}s^2/\kappa},
\label{eq:14}
\end{align} 
with $\kappa=0.804$ and $\mu_{\rm PBE}=0.21951$. By using $E^{\rm x}[n({\bm r})]$ 
in Eq.~(\ref{eq:12}), we can derive $v^{\rm x}_{\sigma}({\bm r})$ the exchange part 
of $v^{\rm xc}_{\sigma} (\bm{r};[n_{\sigma}])$ for a spin-$\sigma$ electron as
\begin{align}
v^{\rm x}_{\sigma}({\bm r})=&\frac{\delta E^{\rm x}[n({\bm r})]}{\delta n({\bm r})}
\biggr |_{n({\bm r})=2n_{\sigma}({\bm r})}=
\epsilon_{\rm x}^{\rm unif}(r_s)\left [\frac{4}{3}F_{\rm x}(s)\right .
\nonumber \\
&\left . -v \frac{\partial F_{\rm x}(s)}{s\partial s}
-\left ( u-\frac{4}{3}s^3 \right ) \frac{\partial}{\partial s}
\left ( \frac{\partial F_{\rm x}(s)}{s\partial s} \right )\right ],
\label{eq:15}
\end{align} 
where $u=u({\bm r})$ and $v=v({\bm r})$ are defined, respectively, as
\begin{align}
u({\bm r})=\frac{\mbox{\boldmath$\nabla$}n({\bm r})
\! \cdot \! \mbox{\boldmath$\nabla$}|\mbox{\boldmath$\nabla$}n({\bm r})|}
{[2k_F({\bm r)}]^3n({\bm r})^2},\ 
v({\bm r})=\frac{\mbox{\boldmath$\nabla$}^2n({\bm r})}
{[2k_F({\bm r)}]^2n({\bm r})}.
\label{eq:16}
\end{align} 

On the other hand, the correlation energy functional $E^{\rm c}[n_{\sigma}({\bm r})]$ 
in Eq.~(\ref{eq:11}) is given by
\begin{align}
E_{c}[n({\bm r})]=\int d{\bm r}
\left [ \epsilon_c^{\rm unif}(r_s,\zeta)+H(r_s,\zeta,t)\right ],
\label{eq:16a}
\end{align} 
where the functional $H(r_s,\zeta,t)$ is defined as
\begin{align}
H(r_s,\zeta,t)\!=\!\gamma \phi^3 \ln 
\left \{ \!1\!+\!\frac{\beta_{\rm MB}}{\gamma}t^2 
\left [\frac{1\!+\!At^2}{1\!+\!At^2\!+\!A^2t^4}\right ]\right \},
\label{eq:16b}
\end{align} 
with introducing $t = (3\pi^2/16)^{1/3}s/\sqrt{r_s}\phi(\zeta)$ and the 
function $\phi(\zeta)$ defined as $\phi(\zeta)=[(1+\zeta)^{1/3}+
(1-\zeta)^{1/3}]/2$. Here $\beta_{\rm MB}=0.066725$~\cite{MB68} and the functional 
$A$ is given as
\begin{align}
A=\frac{\beta_{\rm MB}}{\gamma}
\left \{\exp[-\epsilon_c^{\rm unif}(r_s,\zeta)/\gamma \phi^3]-1\right \}^{-1},
\label{eq:16c}
\end{align} 
with $\gamma=(1-\ln 2)/\pi^2$. By the functional derivative of 
$E^{\rm c}[n_{\sigma}({\bm r})]$ with respect to $n_{\sigma}(\bm{r})$, we obtain 
$v^{\rm c}_{\sigma}({\bm r})$. The concrete form for $v^{\rm c}_{\sigma}({\bm r})$ 
is suppressed here.

In PBEsol, the same forms for the exchange and correlation energy functionals are 
adopted with the replacement of $\mu_{\rm PBE}$ and $\beta_{\rm MB}$ by 
$\mu_{\rm GE}\,(=10/81)$~\cite{AK85} and $0.046$, respectively.

\subsection{DMC}
\label{sec:2-5}

The detailed account of the procedure for DMC is given in Ref.~\cite{TMY15} and thus 
we will not recapitulate it here, but the point is that the only approximation 
involved in DMC is the so-called ``fixed-node approximation''. As explained 
in Ref.~\cite{TMY15}, we consider that unphysical node-position dependent effects 
will be removed by extracting the $N$-independent results, because the node 
positions depend on $N$ in the fixed boundary condition. 

As for the on-top density $n(0)$, the $N$-independent results are found to be 
obtained, if $N$ becomes as large as about 60 for the proton-embedded EG. Futhermore, 
the center of the sphere is very much separated from any postulated node positions, 
implying that $n(0)$ is the physical quantity least affected by the fixed-node 
approximation. For those reasons, it is well expected that DMC provides accurate, 
if not exact, results for $n(0)$. 


\subsection{On-top density in LDA, PBE, and DMC}
\label{sec:2-6}

We have applied DMC to the system decribed by $H$ in Eq.~(\ref{eq:2}) with $Z=1$ 
and, as reported in Ref.~\cite{TMY15}, we have obtained convergent results at 
$N=58$ for $r_s\leq 2.6$ (the TF-screening region) and $N=60$ otherwise (the 
Kondo-screening region). In the stably convergent closed-shell condition, 
the doubly-degenerate $3s$ energy level corresponding to the Kondo singlet state 
is situated just above (below) the Fermi level for low- (high-)$r_s$ systems, 
leading to the difference in $N$ by 2 between the case of $r_s \leq 2.6$ and 
that of $r_s>2.6$. In order to make a direct comparison with those DMC results, 
both LSDA and PBE have been performed in exactly the same situation as for $N$ and 
the boundary condition at each $r_s$. Note that there is no difference between LSDA 
and LDA in the present system, because the ground states are always found to be 
paramagnetic. Thus we will simply write ``LDA'' herafter, even though the actual 
calculations are done in LSDA. 

\begin{figure}[hbt]
  \centering
\resizebox{0.49\textwidth}{!}{%
  \includegraphics{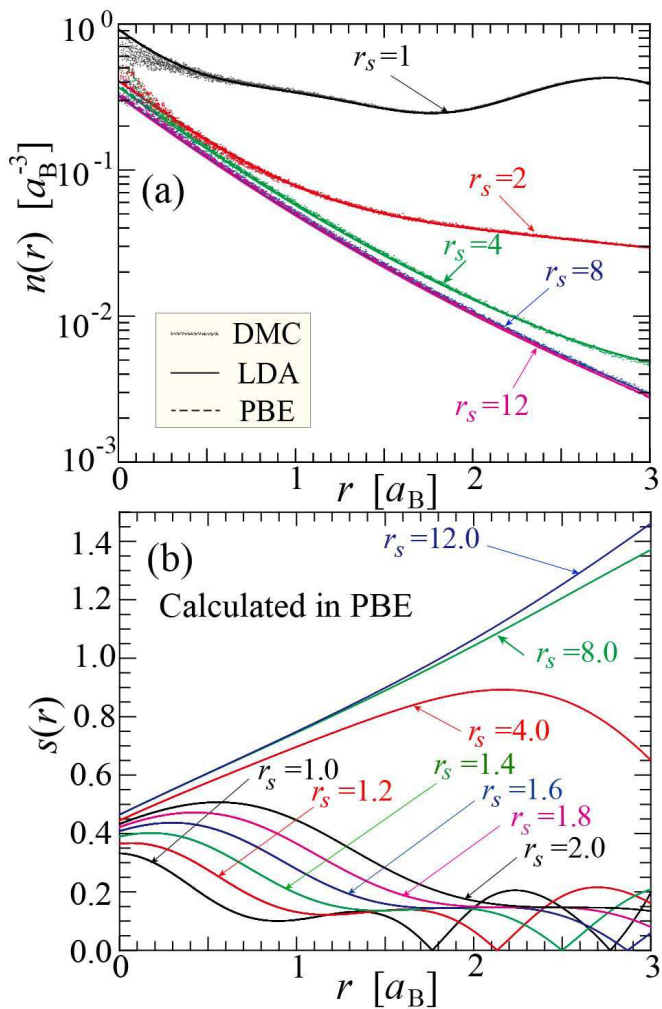}
}
\caption{(a)Density distribution $n({\bm r})$ calculated in LDA, PBE, and DMC for 
the proton-embedded electron-gas sphere with the total electron number $N=58$ for 
$r_s=1$ and 2, and $N=60$ otherwise. (b) The corresponding normalized derivative 
function $s({\bm r})$ in PBE.}
\label{fig:1}       
\end{figure}

The obtained results of $n({\bm r})$ in both LDA and PBE are in good agreement with 
those in DMC, especially for $r$ larger than $2a_{\rm B}$, as seen, for example, 
in Fig.~5 in Ref.~\cite{TMY15}, irrespective of either TF or Kondo region. 
Relatively speaking, for $r$ less than about $a_{\rm B}$, however, there are rather 
large differences among DMC, LDA, and PBE, as shown explicitly in Fig.~\ref{fig:1}(a) 
and the largest deviation occurs at $r=0$. Thus it is important to make a detailed 
quantitative comparison of the on-top density $n(0)$ among those calculation methods 
in order to assess the performance of DFT-based schemes in reference to DMC. 

In view of Eq.~(\ref{eq:1}), $n({\bm r})$ changes linearly with the increase of $r$ 
in semi-log plot, as long as $r$ is less than about $0.3a_{\rm B}$. (In dense 
systems like $r_s$ less than about 2, this critical value for the cusp theorem 
$r_{\rm cusp}$ becomes smaller; it may be safe to take $0.1a_{\rm B}$ for 
$r_{\rm cusp}$ at $r_s=1$.) This linear property in semi-log plots is very useful 
in estimating $n(0)$ in DMC. The results so obtained for $n(0)$ in each scheme are 
given in Table~\ref{tab:1}, from which we find that for $r_s$ less than about 1.66 
(Region I), $n_{\rm DMC}(0)<n_{\rm LDA}(0)<n_{\rm PBE}(0)$, while for larger $r_s$ 
(Region II which includes the TF-Kondo transition point), $n_{\rm DMC}(0)>
n_{\rm PBE}(0)>n_{\rm LDA}(0)$. Note that this interesting crossover point from 
Region I to Region II is situated in the density region in which the solid hydrogen 
and related materials under high pressures are involved, i.e., 
$1.1<r_s<1.7$~\cite{AF13,CHCM16}, making the present assessment relevant and 
important in studying physics of the solid hydrogen in the framework of DFT. 

Incidentally, in Fig.~\ref{fig:1}(b), the results for the normalized derivative 
$s({\bm r})$ in PBE defined in Eq.~(\ref{eq:13}), corresponding to those of 
$n({\bm r})$ in Fig.~\ref{fig:1}(a), are plotted, revealing the interesting fact 
that in Region I, $s({\bm r})$ always stays less than 0.43, but in Region II, 
it beccomes larger than that value. 
It must also be noted that in the Kondo-screening regime in Region II, the behavior 
of $s({\bm r})$ is much different from that in the TF-screening regime, providing 
another piece of evidence for the qualitative difference between those two regimes 
of screening. 

\begin{table}[tbh]
\caption{On-top density $n(0)$ in atomic units for the proton-embedded electron-gas 
sphere with the total electron number $N=58$ for $r_s=1.0-2.6$ and $N=60$ otherwise.}
\label{tab:1}       
  \centering
\begin{tabular}{r|ccc}
\hline\noalign{\smallskip}
$r_s$ & LDA & PBE & DMC \\
\noalign{\smallskip}\hline\noalign{\smallskip}
  $1.0$  &  $0.91886\ $ & $0.92150\ $ & $0.894\pm0.034$ \\
  $1.2$  &  $0.69298\ $ & $0.69702\ $ & $0.674\pm0.023$ \\
  $1.4$  &  $0.56659\ $ & $0.57217\ $ & $0.556\pm0.017$ \\
  $1.6$  &  $0.48990\ $ & $0.49706\ $ & $0.486\pm0.019$ \\
  $1.8$  &  $0.44076\ $ & $0.44946\ $ & $0.450\pm0.014$ \\
  $2.0$  &  $0.40806\ $ & $0.41815\ $ & $0.418\pm0.010$ \\
  $2.2$  &  $0.38570\ $ & $0.39698\ $ & $0.402\pm0.009$ \\
  $2.6$  &  $0.35902\ $ & $0.37198\ $ & $0.382\pm0.008$ \\
  $2.7$  &  $0.45749\ $ & $0.46823\ $ & $0.476\pm0.007$ \\
  $3.0$  &  $0.42502\ $ & $0.43576\ $ & $0.442\pm0.006$ \\
  $4.0$  &  $0.36922\ $ & $0.37950\ $ & $0.388\pm0.006$ \\
  $5.0$  &  $0.34656\ $ & $0.35639\ $ & $0.359\pm0.005$ \\
  $6.0$  &  $0.33550\ $ & $0.34551\ $ & $0.349\pm0.003$ \\
  $7.0$  &  $0.32939\ $ & $0.34027\ $ & $0.347\pm0.003$ \\
  $8.0$  &  $0.32573\ $ & $0.33796\ $ & $0.345\pm0.002$ \\
  $9.0$  &  $0.32339\ $ & $0.33718\ $ & $0.343\pm0.002$ \\
  $10.0$ &  $0.32182\ $ & $0.33711\ $ & $0.342\pm0.001$ \\
  $11.0$ &  $0.32073\ $ & $0.33731\ $ & $0.341\pm0.001$ \\
  $12.0$ &  $0.31995\ $ & $0.33759\ $ & $0.339\pm0.001$ \\
\noalign{\smallskip}\hline
\end{tabular}
\end{table}

\section{Proposal of cusp-corrected PBE}
\label{sec:3}
\subsection{Violation of the cusp theorem in PBE}
\label{sec:3-1}

Confronted with the interesting behavior of the difference between PBE and DMC 
with the increase of $r_s$ in Table~\ref{tab:1}, we have made various trials to 
construct a new xc energy functional in GGA so that $n(0)$ in DMC can be well 
reproduced in the entire range of $r_s$, mostly by just modifying $F_{\rm x}(s)$ 
from the original one in PBE, as is usually the case in most other modifications of 
$E^{\rm xc}[n_{\sigma}({\bm r})]$ from PBE, such as WC~\cite{WC06}. Incidentally, 
there is no problem in Region II; LDA already provides reasonably good $n(0)$ and PBE 
improves much on it, but it is by no means easy to obtain $n(0)$ in similar accuracy 
in Region I. Thus we will focus on that region in the following. 

In Region I, $s({\bm r})$ is less than 0.43 and thus we need some new insight into 
$E^{\rm xc}[n(\bm r)]$ in this small-$s$ range. In pursuit of the new ingredient 
needed for improving on the PBE energy functional, we have paid attention to the cusp 
theorem; as mentioned in Sect.~\ref{sec:1}, the cusp behavior in Eq.~(\ref{eq:1}) is 
correctly reproduced in LDA, but it is usually not the case in GGA due to the 
appearance of a singular term $-\delta Z/|{\bm r}|$ in the exchange-correlation 
potential $v^{\rm xc}({\bm r})$ near the nucleus, in addition to the external 
singular term $-Z/|{\bm r}|$. In the presence of this additional singular term, 
the cusp behavior is not determined by $Z$ but $Z+\delta Z$, leading to the relative 
error in proportion to $\delta Z/Z$.

With the use of Eqs.~(\ref{eq:1}), (\ref{eq:13}), and (\ref{eq:16}), we find that for 
$r \approx 0$, $s({\bm r})$, $u({\bm r})$, and $v({\bm r})$ behave, respectively, as 
\begin{align}
s({\bm r}) \approx s_c, \quad 
u({\bm r}) \approx s_c^3, \quad
v({\bm r}) \approx s_c^2-\frac{s_c^2}{Zr},
\label{eq:17a}
\end{align} 
with $s_c\equiv Z/[3\pi^2n(0)]^{1/3}$. Thus for $r \approx 0$, the singular 
contribution to $v^{\rm x}_{\sigma}({\bm r})$ in Eq.~(\ref{eq:15}) comes only 
from the term in proportion to $v({\bm r})$. More explicitly, the singular term can 
be written as
\begin{align}
-\epsilon_{\rm x}^{\rm unif}(r_s^c) \left ( -\frac{s_c^2}{Zr}\right )
\frac{\partial F_{\rm x}(s_c)}{s_c\partial s_c}
=-\!\left(\!\frac{3}{4\pi}\frac{\partial F_{\rm x}(s_c)}{\partial s_c}\!\right )
\frac{1}{r},
\label{eq:17b}
\end{align} 
with $r_s^c\equiv [3/4\pi n(0)]^{1/3}$. Similarly, the singular term in 
$v^{\rm c}_{\sigma}({\bm r})$ is written in the form of Eq.~(\ref{eq:17b}) with the 
replacement of $F_{\rm x}(s_c)$ by $F_{\rm c}(s_c)$, defined as
\begin{align}
F_{\rm c}(s_c)=\frac{\epsilon_{\rm c}^{\rm unif}(r_s^c,0)+H(r_s^c,0,t_c)}
{\epsilon_{\rm x}^{\rm unif}(r_s^c)}
\label{eq:17c}
\end{align} 
with $t_c=(3\pi^2/16)^{1/3}s_c/\sqrt{r_s^c}$. Then $\delta Z$ is given by 
\begin{align}
\delta Z = \frac{3}{4\pi}\frac{\partial}{\partial s}
\left ( F_{\rm x}\!+\!F_c\right )
= \frac{3}{2\pi}s\left (\frac{\partial F_{\rm x}}{\partial s^2}+
\frac{\partial F_{\rm c}}{\partial s^2} \right),
\label{eq:18}
\end{align} 
evaluated at the cusp position $r=0$ with $r_s=r_s^c$, $s=s_c$, and $\zeta=0$. 

For the case of $Z\gg 1$, $n(0)$ is well approximated by either $Z^3/\pi$ in the 
strong-correlation limit or $2Z^3/\pi$ in the weak--correlation limit. Then we 
obtain $s_c$ and $r_s^c$, respectively, as either $(3\pi)^{-1/3}\!\approx \!0.473$ 
and $(3/4)^{1/3}Z^{-1}$ or $(6\pi)^{-1/3}\!\approx \!0.376$ and $(3/8)^{1/3}Z^{-1}$ 
in each limit, implying that $s_c$ is in the range $(0.376,0.473)$ and $r_s^c\ll 1$. 
However, not only in the present atom-embedded EG but also in atoms, 
molecules, and solids in which the condition of $Z\gg 1$ is not always satisfied, 
$s_c$ varies in the range from 0.32 to 0.473, still a relatively small range of 
$s$ around 0.4. 

\begin{figure}[htb]
  \centering
\resizebox{0.47\textwidth}{!}{%
  \includegraphics{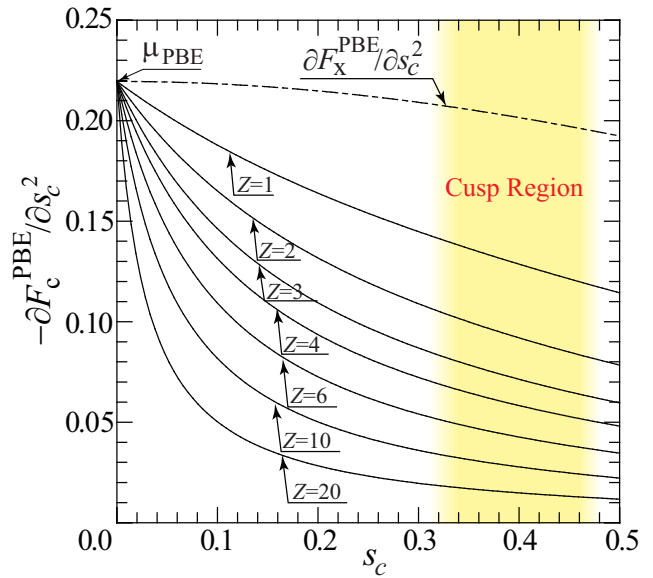}
}
\caption{``Target function'' or the partial derivative of $-F_{\rm c}$ with respect 
to $s_c^2$ in PBE plotted as a function of $s_c$ for various $Z$ with 
$r_s\!=\!(9\pi/4)^{1/3}s_c/Z$. For comparison, $\partial F_{\rm x}/\partial s_c^2$ 
in PBE is also plotted by the dotted-dashed curve. In actual cusp positions, 
the value of $s_c$ is in the range $0.32-0.473$.}
\label{fig:2}       
\end{figure}

Now, let us take $F_{\rm c}$ as $F_{\rm c}^{\rm PBE}$ the one given in PBE. 
Then the second component in Eq.~(\ref{eq:18}) or the function 
$\partial F_{\rm c}^{\rm PBE}/\partial s_c^2$ is concretely known 
as a function of $s_c$ with $r_s^c$ set equal to $(9\pi/4)^{1/3}s_c/Z$ for each 
$Z$. In Fig.~\ref{fig:2}, this function (or actually its negative, $-\partial 
F_{\rm c}^{\rm PBE}/\partial s_c^2$) is plotted as a function of $s_c$, from which 
we see that if $s_c$ were zero (or at least very small), the cusp condition would 
be (almost) fulfilled in PBE, because $\partial F_{\rm c}^{\rm PBE}/\partial s_c^2$ 
(which is eqaul to $-\mu_{\rm PBE}$ at $s_c=0$, irrespective of $Z$) is cancelled 
by $\partial F_{\rm x}^{\rm PBE}/\partial s_c^2=\mu_{\rm PBE}/(1\!+\!\mu_{\rm PBE}
s_c^2/\kappa)^2$. In fact, in the original PBE, $\mu_{\rm PBE}$ is so determined as 
to satisfy this condition at $s_c=0$, a relation to intimately connect 
$F_{\rm x}^{\rm PBE}$ with $F_{\rm c}^{\rm PBE}$. In the actual cusp region in which 
$s_c$ is about 0.4, however, $\delta Z$ is not small enough and thus the cusp theorem 
is voilated in PBE; the relative error $\delta Z/Z$ is about 1.8\% and 1.2\% for $Z=1$ 
and $2$, respectively, and less than 1\% for $Z \ge 3$. 

The above observation inspires us that if we can modify $F_{\rm x}(s)$ so as to 
cancel $\partial F_{\rm x}(s)/\partial s^2$ with $\partial F_{\rm c}^{\rm PBE}
/\partial s^2$ for $s$ not at a single point of $s=0$ but in the entire cusp 
region of $(0.32,0.473)$, we can always make $\delta Z$ vanish at the cusp point 
(and thus impose the cusp theorem), whatever value for $s_c$ is determined in the 
self-consistent calculation of $n({\bm r})$. This constitutes the main idea of this 
paper. Of course, because $-\partial F_{\rm c}^{\rm PBE}/\partial s^2$, which will 
be called ``the target function'' hereafter, depends on $Z$, we have to treat 
its $Z$ dependence appropriately, but for the time being, we take the target 
function at $Z=1$. Then, for other values of $Z$ the cusp theorem will be violated, 
but in this case the relative error $\delta Z/Z$ becomes much smaller than that in 
PBE; at the most, it is about 0.36\% for $Z=2$ or $3$. 

A formally better scheme to impose the constraint due to the cusp theorem for any 
$Z$ will be mentioned in Sect.~\ref{sec:4}. As for the choice of $F_{\rm c}$, we 
have examined the case of $F_{\rm c}$ with $\beta_{\rm MB}$ in Eqs.~(\ref{eq:16b}) 
and (\ref{eq:16c}) replaced by either 0.046 as in PBEsol or the more refined 
$r_s$-dependent one, $\beta(r_s)$, expressed as~\cite{PRCCS09,HL86}
\begin{align}
\beta(r_s)=\beta_{\rm MB}\,\frac{1+0.1r_s}{1+0.1778r_s},
\label{eq:32a}
\end{align} 
but we find that no appreciable difference is seen in the final results for $n(0)$. 

\subsection{Exchange Energy Functional in ccPBE}
\label{sec:3-2}

In order to construct $F_{\rm x}(s)$ in accordance with the above-mentioned idea 
to fulfill the cusp theorem under the assumption that the correlation energy 
functional is set equal to $F_{\rm c}^{\rm PBE}$, we have examined a variety of 
possible forms to arrive at the following $F_{\rm x}(s)$ which is given as the sum 
of three terms:
\begin{align}
F_{\rm x}(s)=F_0(s)+F_1(s)+F_2(s),
\label{eq:19}
\end{align} 
where $F_0(s)$ is basically the one only slightly modified from the original form 
in PBE as
\begin{align}
F_0(s)=A_0+A_1-\frac{A_1}{1+\mu(p)\,p/A_1},
\label{eq:20}
\end{align} 
where $p\equiv s^2$ and $\mu(p)$ is assumed to be
\begin{align}
\mu(p)=\mu_1+(\mu_0-\mu_1)\exp(-p/s_0^2).
\label{eq:21}
\end{align} 
The function $F_1(s)$ is assumed to be
\begin{align}
F_1(s)=B_0\exp(-p^2/s_1^4),
\label{eq:22}
\end{align} 
in order to satisfy the exact gradient expansion (GE) of $F_{\rm x}(s)$ 
in the limit of $s \to 0$, known as~\cite{SvB96}
\begin{align}
F_{\rm x}=1\!+\!\mu_{\rm GE}\,p\!+\!\frac{146}{2015}v^2
\!-\!\frac{73}{405}p\,v\!+\!Dp^2\!+\!O(\mbox{\boldmath$\nabla$}^6),
\label{eq:23}
\end{align} 
where $v$ is defined in Eq.~(\ref{eq:16}) and the coefficient $D$ vanishes according 
to the best numerical estimate. The function $F_2(s)$ is so introduced as to impose 
the constraint due to the cusp theorem; namely, $\partial F_{\rm x}(s)/\partial s^2$ 
is set equal to $-\partial F_{\rm c}^{\rm PBE}/\partial s^2$ with $Z=1$ for $s$ in 
the range $(0.32,0.473)$. The actual procedure is to begin with the assumption of 
$F_2(s)$ in the form of
\begin{align}
F_2(s)=&\frac{p^2}{s_2^4}
\left[C_0+\sum_{i=1}^6 C_i \left(\frac{p}{s_2^2} \right )^i \right ]\exp(-p/s_2^2).
\label{eq:24}
\end{align} 
Then, under given values for $C_0$ and $s_2$, we determine the six coefficients, 
$C_1, \cdots, C_6$, so as to satisfy the above-mentioned condition for fulfilling 
the cusp theorem. 

\begin{table}[tbh]
\caption{Set of parameters to specify $F_{\rm x}(s)$ in ccPBE. In order to satisfy 
the cusp condition, three target functions corresponding to $Z=1$, 2, and 3 are 
considered. Note that only the parameters $C_0,C_1,\cdots,C_6$, and $s_2$ depend 
on $Z$.}
\label{tab:2}       
  \centering
\begin{tabular}{c|lll}
\hline\noalign{\smallskip}
 & $Z=1$ & $Z=2$ & $Z=3$ \\
\noalign{\smallskip}\hline\noalign{\smallskip}
  $A_0$   &  $\ 1.036$         & $\ 1.036$           & $\ 1.036$           \\
  $A_1$   &  $\ 0.768$         & $\ 0.768$           & $\ 0.768$           \\
  $\mu_0$ &  $\ 0.12345679$    & $\ 0.12345679$      & $\ 0.12345679$      \\
  $\mu_1$ &  $\ 0.13170898$    & $\ 0.13170898$      & $\ 0.13170898$      \\
  $s_0$   &  $\ 1.20$          & $\ 1.20$            & $\ 1.20$            \\
  $B_0$   &  -$0.036$          & -$0.036$            & -$0.036$            \\
  $s_1$   &  $\ 0.180$         & $\ 0.180$           & $\ 0.180$           \\
  $C_0$   &  -$0.006933655 $   & -$0.007332614 $     & -$0.007538429 $     \\
  $C_1$   &  -$0.011363996 $   & -$0.036651747 $     & -$0.038238364 $     \\
  $C_2$   &  $\ 0.010829969$   & $\ 0.039997327$     & $\ 0.046200149$     \\
  $C_3$   &  -$0.003780562 $   & -$0.014587676 $     & -$0.017395384 $     \\
  $C_4$   &  $\ 0.000643956$   & $\ 0.002663050$     & $\ 0.003272557$     \\
  $C_5$   &  -$0.0000554048$   & -$0.0002416611$     & -$0.0003038872$     \\
  $C_6$   &  $\ 0.00000197693$ & $\ 0.00000954069$   & $\ 0.00001237041$   \\
  $s_2$   &  $\ 0.142        $ & $\ 0.144        $   & $\ 0.145        $   \\
\noalign{\smallskip}\hline
\end{tabular}
\end{table}

There are still nine papameters, $s_0$, $s_1$, $s_2$, $A_0$, $A_1$, $\mu_0$, 
$\mu_1$, $B_0$, and $C_0$, to be fixed, but they cannot be chosen independently; 
there are four important constraints; in the limit of $s \to \infty$, there is the 
Lieb-Oxford upper bound $1+\kappa$ with $\kappa=0.804$ for $F_{\rm x}(s)$~\cite{LO81}, 
leading to the condition of 
\begin{align}
\label{eq:25}
\lim_{s\to \infty} F_{\rm x}(s)=A_0+A_1=1+\kappa.
\end{align} 
In the limit of $s \to 0$, we should respect Eq.~(\ref{eq:23}), but because 
the functional $F_{\rm x}$ in GGA is assumed to be a function of a single variable 
$s$, we need to derive an approximate expression for $v$ in terms of $s$ in 
order to make use of Eq.~(\ref{eq:23}). As in Eq.~(\ref{eq:17a}), by the use of the 
definitions of $s$ and $v$ in Eqs.~(\ref{eq:13}) and (\ref{eq:16}), respectively, 
and the behavior of $n({\bm r})$ in Eq.~(\ref{eq:1}) near the nucleus at which the 
electron density varies most rapidly, we obtain
\begin{align}
s=\frac{Z}{k_F}\quad {\rm and}\quad 
v=\frac{Z^2}{k_F^2}\left (1-\frac{1}{Zr}\right).
\label{eq:26}
\end{align} 
Then, $v$ is approximately given by $v=v_0 s^2$ with a coefficient $v_0$ 
which is calculated by taking the average of $1/Zr$ by the weight of $n(0)
\exp(-2Zr)$ in the range of $0\leq r \leq r_{\rm cusp}$ with $r_{\rm cusp}$ 
which is the critical value for $r$ satisfying the cusp condition, as introduced in 
Sect.~\ref{sec:2-6}. More specifically, $v_0$ is calculated as
\begin{align}
v_0&=1-\left \langle \frac{1}{Zr} \right \rangle
=1 - \frac{\int d{\bm r} \exp(-2Zr)/Zr}{\int d{\bm r} \exp(-2Zr)}
\nonumber \\
&=-\frac{\rho_0^2/2}{\exp(\rho_0)-1-\rho_0-\rho_0^2/2},
\label{eq:27}
\end{align} 
with $\rho_0=2Zr_{\rm cusp}$. Because $r_{\rm cusp}$ is about 0.1 or larger and 
the case of $Z=1$ is considered here, we take $\rho_0$ as 0.20 tentatively in the 
following. Then, we obtain the small-$s$ expansion of 
$F_{\rm x}(s)$ in the following way:
\begin{align}
F_{\rm x}(s) =1+\mu_{\rm GE}\,p+\mu_4 \,p^2,
\label{eq:28}
\end{align} 
with $\mu_4=(146/2015)v_0^2-(73/405)v_0 \approx 17.22612$. 

In accordance with the small-$p$ expansion in Eq.~(\ref{eq:28}), terms at each 
order, $O(p^0)$, $O(p)$, or $O(p^2)$, should satisfy
\begin{align}
\label{eq:29}
F_{\rm x}(0)&=1=A_0+B_0,\\
\label{eq:30}
\frac{\partial F_{\rm x}(0)}{\partial p} &= \mu_{\rm GE}=\mu_0,\\
\label{eq:31}
\frac{1}{2}\frac{\partial^2 F_{\rm x}(0)}{\partial^2 p} &= \mu_4
=\frac{\mu_1-\mu_0}{s_0^2}-\frac{\mu_0^2}{A_1}-\frac{B_0}{s_1^4}+\frac{C_0}{s_2^4},
\end{align} 
respectively. By use of Eqs.~(\ref{eq:25}), (\ref{eq:29})-(\ref{eq:31}), the 
parameters, $A_0$, $A_1$, $\mu_0$, and $C_0$ can be determined under given values for 
the rest of the parameters. 

By comparing the calculated results for the single-proton embedded electron-gas 
sphere in ccPBE with those in DMC, we can determine an appropriate set of parameters 
providing sufficiently good results. The parameter set so obtained is given in 
Table~\ref{tab:2} in which another set of parameters fulfilling the cusp condition 
for $Z=2$ and $3$ are also added. Note that the parameters depending on $Z$ are only 
those concerned with $F_2(s)$. 

\begin{figure}[htb]
  \centering
\resizebox{0.49\textwidth}{!}{%
  \includegraphics{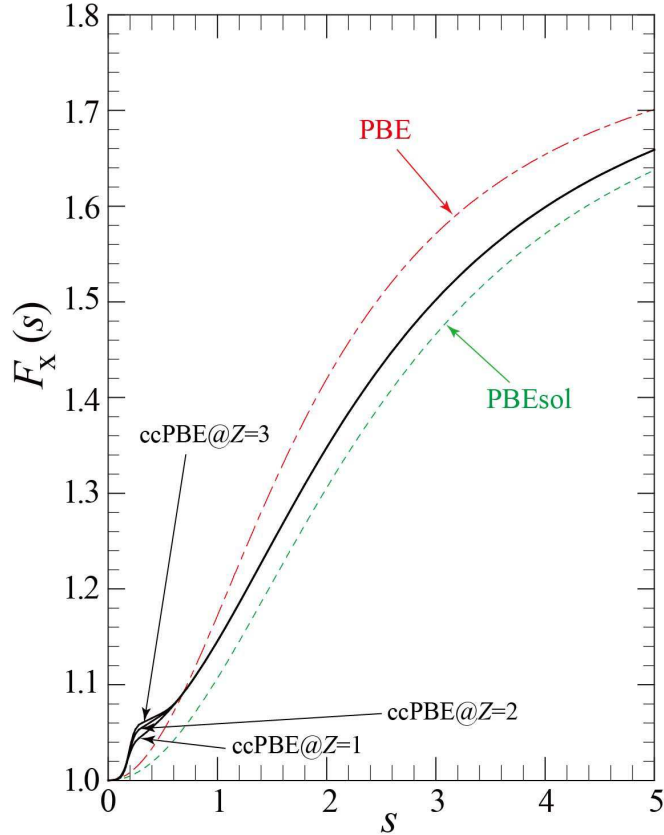}
}
\caption{$F_{\rm x}(s)$ in ccPBE determined in reference to three different target 
functions with $Z=1$, 2, and 3. For comparison, we also plot $F_{\rm x}(s)$ in 
both PBE and PBEsol by dotted dashed and dotted curves.}
\label{fig:3}       
\end{figure}

\begin{figure}[hbt]
  \centering
\resizebox{0.49\textwidth}{!}{%
  \includegraphics{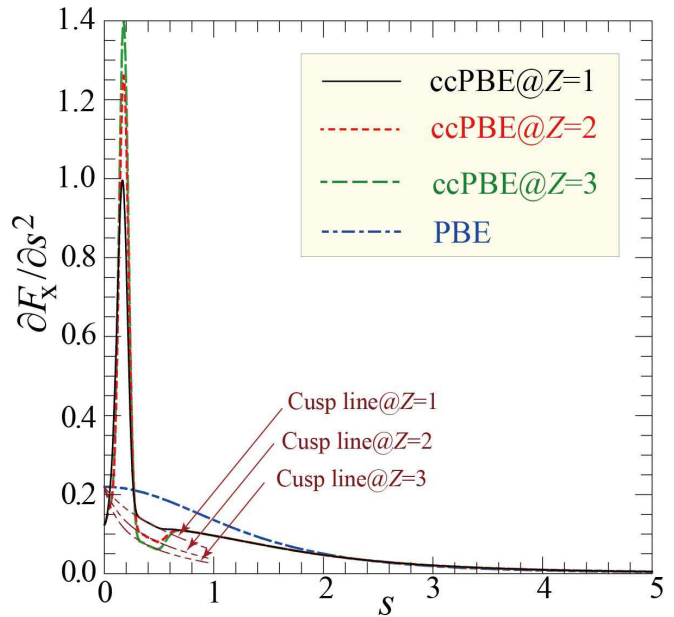}
}
\caption{$\partial F_{\rm x}/\partial s^2$ corresponding to $F_{\rm x}(s)$ in ccPBE 
in Fig.~\ref{fig:3}. It coincides with the target function (or the cusp line) at 
each $Z$, plotted by the double-dotted-dashed curve, for $s$ in the range 
$0.32-0.473$.}
\label{fig:4}       
\end{figure}

With those parameters, we can concretely give $F_{\rm x}(s)$ in ccPBE and its 
derivative $\partial F_{\rm x}(s)/\partial s^2$, both of which are plotted 
in Figs.~\ref{fig:3} and \ref{fig:4}, respectively. We have plotted $F_{\rm x}(s)$ 
for three cases of the target functions with $Z=1$, 2, and 3, but its $Z$-dependence 
is found to be weak. Compared with $F_{\rm x}(s)$ in PBE and PBEsol, 
$F_{\rm x}(s)$ in ccPBE is enhanced much and has a characteristic structure 
for $s<0.6$ but it increases smoothly for $s>0.6$ and its actual value comes to 
the middle of PBE and PBEsol. 

As for $\partial F_{\rm x}(s)/\partial s^2$, its $Z$-dependence is much stronger than 
that for $F_{\rm x}(s)$. The first sharp peak at $s \approx 0.2$ is found to be 
important to control the actual values of $n(0)$ for $r_s<2$. (See the dependence 
of the peak structure on $Z$ in Fig.~\ref{fig:4} and the change of $n(0)$ with 
$Z$ in Table~\ref{tab:3}.) This structure appears by reconciliation of the two 
constraints, one from the gradient expansion, Eq.~(\ref{eq:23}) or Eq.~(\ref{eq:28}), 
and the other from the cusp condition which is explicitly shown by ``the cusp lines'' 
in Fig.~\ref{fig:4}. 

\subsection{On-top density in ccPBE}
\label{sec:3-3}

In ccPBE, we have successfully applied to the proton-embedded electron-gas sphere 
and obtained very good results for $n(0)$ in the entire range $1-12$, as given in 
Table~\ref{tab:3}. The TF-Kondo transition is seen by the jump in $n(0)$ at 
$r_s \approx 2.6$. Accuracy of the results in ccPBE is estimated by the relative 
error with respect to the DMC data, given in \%. Note that {\it nineteen} 
independent data in DMC are reproduced very well by the appropriate choice of 
only {\it five} free parameters in ccPBE. 

For the target function with $Z=1$, which agrees with the atomic number of proton, 
the errors are at most about 1\% but mostly much less than 1\%. If we employ the 
target function with $Z=2$, which is twice as large as the atomic number of proton, 
in determining $F_2(s)$ in Eq.~(\ref{eq:19}) or Eq.~(\ref{eq:24}), the errors are 
about several \%, which may be said to be much larger than the case of $Z=1$, 
but at the same time it may be said to be still small enough compared to the case 
of the original PBE. This better perfomance may be said to be due to the much smaller 
error in $\delta Z/Z$ in ccPBE, even though we do not employ the target function 
with the correct value of $Z$. Incidentally, if we calculate $n(0)$ in ccPBE with 
using the target function with $Z=3$, the relative errors are found to be still 
not large, ranging from -1.3\% to 7.0\%, about twice as large as those in the case 
of $Z=2$. 

\begin{table}[tbh]
\caption{On-top density $n(0)$ in ccPBE for the proton-embedded electron-gas 
sphere with the total electron number $N=58$ for $r_s=1.0-2.6$ and $N=60$ otherwise. 
The parameter sets are used for the target function with either $Z=1$ or $Z=2$. 
The errors are given as the relative ones in \% with respect to the DMC results.}
\label{tab:3}       
  \centering
\begin{tabular}{r|cc|cc}
\hline\noalign{\smallskip}
 & ccPBE      &  & ccPBE       &  \\
$r_s$ & @$Z\!=\!1$ & Error(\%)   & @$Z\!=\!2$  & Error(\%) \\
\noalign{\smallskip}\hline\noalign{\smallskip}
  $1.0$  &  $0.89459$ & $\ 0.07$ &  $0.88545\ $ & -$0.96$  \\
  $1.2$  &  $0.67670$ & $\ 0.40$ &  $0.66729\ $ & -$1.00$  \\
  $1.4$  &  $0.55895$ & $\ 0.53$ &  $0.54989\ $ & -$1.10$  \\
  $1.6$  &  $0.48948$ & $\ 0.72$ &  $0.48177\ $ & -$0.87$  \\
  $1.8$  &  $0.44527$ & -$1.05$  &  $0.44013\ $ & -$2.19$  \\
  $2.0$  &  $0.41785$ & -$0.04$  &  $0.42272\ $ & $\ 1.13$  \\
  $2.2$  &  $0.40152$ & -$0.12$  &  $0.41786\ $ & $\ 3.94$  \\
  $2.6$  &  $0.38202$ & $\ 0.01$ &  $0.40267\ $ & $\ 5.41$  \\
  $2.7$  &  $0.47532$ & -$0.14$  &  $0.48558\ $ & $\ 2.01$  \\
  $3.0$  &  $0.44295$ & $\ 0.22$ &  $0.45574\ $ & $\ 3.11$  \\
  $4.0$  &  $0.38559$ & -$0.62$  &  $0.40013\ $ & $\ 3.13$  \\
  $5.0$  &  $0.36097$ & $\ 0.55$ &  $0.37486\ $ & $\ 4.42$  \\
  $6.0$  &  $0.34917$ & $\ 0.05$ &  $0.36299\ $ & $\ 4.01$  \\
  $7.0$  &  $0.34335$ & -$1.05$  &  $0.35682\ $ & $\ 2.83$  \\
  $8.0$  &  $0.34100$ & -$1.16$  &  $0.35406\ $ & $\ 2.63$  \\
  $9.0$  &  $0.34064$ & -$0.69$  &  $0.35361\ $ & $\ 3.10$  \\
  $10.0$ &  $0.34105$ & -$0.28$  &  $0.35394\ $ & $\ 3.49$  \\
  $11.0$ &  $0.34168$ & $\ 0.20$ &  $0.35467\ $ & $\ 4.01$  \\
  $12.0$ &  $0.34221$ & $\ 0.95$ &  $0.35525\ $ & $\ 4.79$  \\
\noalign{\smallskip}\hline
\end{tabular}
\end{table}

\section{Summary and Discussion}
\label{sec:4}

By imposing the consraint originating from the cusp theorem on the PBE scheme in 
GGA to DFT, we have sucessfully constructed a new exchange-correlation energy 
functional, referred to as ccPBE (cusp-corrected PBE), and accurately reproduced 
the DMC data on the on-top electron density $n(0)$ in the proton-embedded electron 
gas with the density parameter $r_s$ in the range $1-12$. 

Five comments are in order: (i) Among fifteen parameters in the definition of 
$F_{\rm x}(s)$ in Eq.~(\ref{eq:19}), only five parameters, namely, $s_0$, $s_1$, 
$s_2$, $\mu_1$, and $B_0$, can be chosen freely and independently of various 
constraints. After a rather extensive search for appropriate values for them, 
we come to notice that the adequate ranges for $s_1$ and $s_2$ are limited by the 
cusp region $(0.32,0.473)$ in $s$-variable space and probably the best values 
for them are those in Table~\ref{tab:2}. In this sense, $F_{\rm x}(s)$ for $s$ 
in the range $s<0.473$ is almost completely determined nonempirically by both 
the exact gradient expansion and the cusp theorem. As for other parameters, namely, 
$s_0$, $\mu_1$, and $B_0$ having strong influence on $F_{\rm x}(s)$ for $s>0.473$, 
it is still not certain whether the set of those values in Table~\ref{tab:2} are 
best or not. A better set of those parameters might be found in the future. 

(ii) As related to the above point, it might be considered that $\rho_0$ in 
Eq.~(\ref{eq:27}) is another independent and important parameter, but it does not 
seem to be the case, because even if $\rho_0=1.0$ is chosen instead of $\rho_0=0.2$ 
and consequently much different values for $C_i$ are used to define $F_2(s)$, the 
self-consistently determined results for $n({\bm r})$ do not change much, 
indicating that we may choose any value of $\rho_0$ as long as it is in the 
physically appropriate range $0.2-1.0$. 

(iii) As for the choice of $Z$ in determining the target function, it is perfectly 
reasonable to choose $Z=1$ for the problems on hydrogen and the parameter set at 
$Z=1$ can be applied as it is to the solid hydrgen under high pressures. Even for 
the case of other values of $Z$, we might say that ccPBE with the parameter set 
at $Z=1$ may provide better results than PBE, but this needs to be confirmed 
by a comprehensive test of ccPBE for a wide class of real materials in the future.
This test will also contribute much to the choice of best appropriate values for 
the parameters $s_0$, $\mu_1$, and $B_0$.

(iv) From a fundamental point of view, the xc functional should be universal 
and must be determined only by the electron density $n({\bm r})$ itself. Thus one 
may argue that the $Z$-dependent xc functional cannot be acceptable from the basic 
principles of DFT. In order to overcome this criticism, we may propose the following 
amendment: Among three terms in Eq.~(\ref{eq:19}), only $F_2(s)$ depends on $Z$ 
through the $Z$-dependence in $C_0$, $\cdots$, $C_6$, and $s_2$. Then, let us rewrite 
Eq.~(\ref{eq:24}) as
\begin{align}
F_2(s,Z)=&\frac{p^2}{s_2(Z)^4}
\left[C_0(Z)+\sum_{i=1}^6 C_i(Z) \left(\frac{p}{s_2^2} \right )^i \right ]
\nonumber \\
&\times \exp(-p/s_2(Z)^2).
\label{eq:32}
\end{align} 
Now, since the term $F_2(s,Z)$ becomes important only in the cusp region at which 
the relation of $Z\!=\!(9\pi/4)^{1/3}s/r_s$ holds, we use its relation to 
introduce the $r_s$-dependent functional $F_{\rm x}(s,r_s)$ as
\begin{align}
F_{\rm x}(s,r_s)=F_0(s)+F_1(s)+F_2(s,(9\pi/4)^{1/3}s/r_s),
\label{eq:33}
\end{align} 
instead of $F_{\rm x}(s)$ in Eq.~(\ref{eq:19}). This functional $F_{\rm x}(s,r_s)$ 
satisfies the basic principles of DFT and at the same time the cusp theorem will be 
satisfied for any $Z$. Note that once we consider the $r_s$-dependence in 
$F_{\rm x}$, there are additional terms in $v^{\rm x}_{\sigma}({\bm r})$ in 
Eq.~(\ref{eq:15}) and due changes must be made in the sebsequent calculations, 
including the determination of the coefficients $C_i$ and their $Z$-dependence. 
All those tasks concerning this amendment must be done before implementing a 
comprehensive test of ccPBE. Those works are left for the future. 

(v) It is argued that the cusp theorem is satisfied in meta-GGA~\cite{SSTP04,PRTSSC05}. 
Then the DMC data in Table~\ref{tab:1} provide a good testing ground for 
meta-GGA. In particular, it would be interesting to see which is the predominat 
scheme among several proposed ones~\cite{TPSS03,PRCCS09,SRP15,ZT06,AK13,TM16} 
in reference to the DMC data.

%
%

\end{document}